\documentclass{cpbtex}
\usepackage{graphicx}

\newcommand{\change}[1]{\textcolor{black}{#1}}

\begin{document}

\title{Simulation of microswimmer hydrodynamics \\ with multiparticle collision dynamics \thanks{This project acknowledges funding from the Austrian Science Fund (FWF) through a Lise-Meitner Fellowship (Grant No M 2458-N36).}}


\author{Andreas Z\"ottl\thanks{Corresponding author. E-mail:~andreas.zoettl@tuwien.ac.at} \\
{Institute for Theoretical Physics, TU Wien, Wiedner Hauptstra{\ss}e 8-10, Wien, Austria}}    


\date{\today}
\maketitle

\begin{abstract}
  In this review  we discuss the recent progress in the simulation of soft active matter systems and in particular the hydrodynamics of microswimmers
  using the method of multiparticle collision dynamics, which
  solves the hydrodynamic flows around active objects on a coarse-grained level. We first present a brief overview of the basic simulation method and the coupling between microswimmers and fluid. \change{We then}  review the current \change{achievements} in simulating flexible and rigid microswimmers using multiparticle collision dynamics, and \change{briefly} conclude and discuss possible future directions.
\end{abstract}



\section{Introduction}
The rapidly growing field of active matter physics describes the self-assembly, structure, and dynamics of various biological and \change{artificial} systems  consisting of active elements out of equilibrium \cite{Marchetti2013}.
Examples range from the collective dynamics of swimming fish or flying birds \cite{Vicsek2012}, down to the autonomous self-propulsion of biological microswimmers \cite{Purcell1977,Lauga2009a,Elgeti2015,Zoettl2016}, or intracellular active processes  \cite{Prost2015}.
Microswimmers are able to propel themselves either by a non-reciprocal deformation of their cell body or by self-generating near-surface flows as it is the case for many artificial microswimmers \cite{Zoettl2016}.

When microswimmers move in a fluid they self-generate different fluid flows.
Since neutrally buoyant swimmers can move without applying external forces, the leading order flow field  is typically a force dipole field.
The specific body deformation or propulsion mechanism then determines the sign of the force dipole strength $p$.
So-called pushers typically have their propelling \change{apparatus} at the back and push fluid backwards ($p>0$), while so-called pullers have it in front of their body ($p<0$).
For some microswimmers the force dipole contribution can be weak, for example for some spherical artificial swimmers, and the flow field is then well approximated by a source doublet field \cite{Zoettl2016}.

Recently coarse-grained simulation techniques became a powerful tool to simulate the fluctuating hydrodynamics of many different passive and active soft matter systems on the nano- and micron scale.
The most important examples are the Lattice Boltzmann (LB) method \cite{Dunweg2009}, Dissipative Particle Dynamics (DPD) \cite{Espanol2017}, and Multiparticle Collision Dynamics (MPCD) \cite{Gompper2009,Kapral2008}.
Simulating active system at the micron scale using MPCD is a relatively new field of research
used for many different applications at the nano- and micron scale
in the last 15 years.
In this review we discuss the advances made in understanding the dynamics of microswimmers using MPCD.

\section{Multiparticle collision dynamics}
\subsection{Simulation of the fluid}
The method of MPCD
was originally  proposed by Malevanets and Kapral and termed \textit{stochastic rotation dynamics} (SRD) \cite{Malevanets1999,Malevanets2000b}.
It solves the Navier-Stokes equations on a coarse-grained level
and naturally includes thermal fluctuations \cite{Kapral2008,Gompper2009}.
This is achieved by modeling the fluid by $N_p$ pointlike, effective fluid particles at positions $\mathbf{r}$ and velocities  $\mathbf{v}$ and each of mass $m_0$ and kept at temperature $T_0$.

At the beginning of the simulation the fluid particles are randomly placed in the simulation box and their velocities are drawn from a Gaussian distribution with standard deviation 
$\sqrt{k_BT_0/m_0}$ with $k_B$ the Boltzmann constant.
Then they perform alternating \textit{streaming} and  \textit{collision} steps which are sufficient to solve the Navier-Stokes equations on a coarse-grained level
because of momentum conservation in the collision step.

\begin{figure}[bth]
\begin{center}
\includegraphics[width=7.4cm]{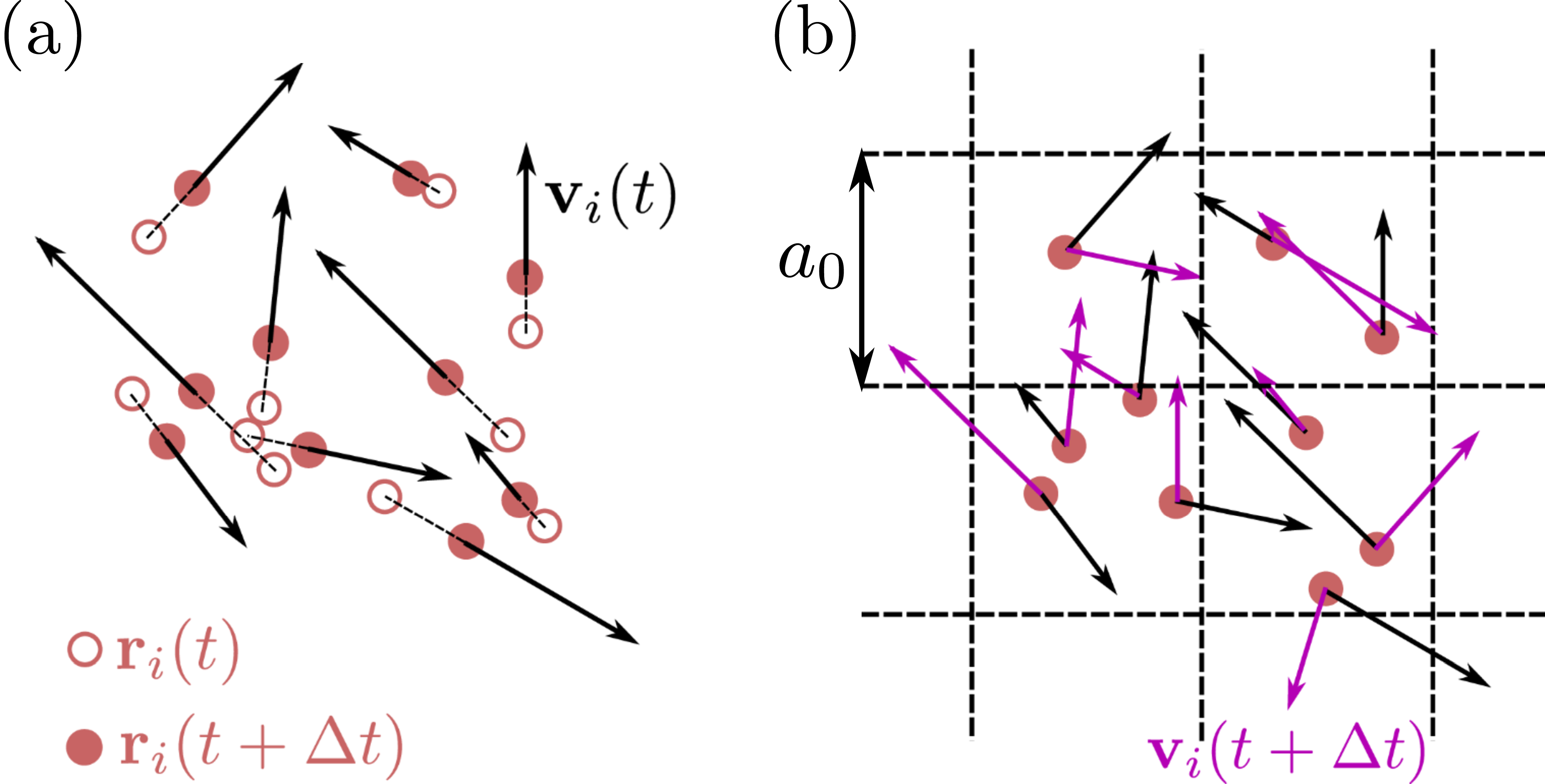}
\end{center}
\caption{2D Sketch of the MPCD method (from Ref.~\cite{Zoettl2016}). (a) In the streaming step the fluid particles move ballistically.
(b) In the collision step the fluid particles are sorted into cubic cells and exchange momentum with all other particles in the cell
while keeping the local total momentum fixed. }
\label{Fig:1}
\end{figure}

Figure~\ref{Fig:1} shows a 2D sketch of the streaming and collision step.
In the streaming step the  particles move ballistically for a time $\Delta t$ and the positions are updated to (Fig.~\ref{Fig:1}(a))
\begin{equation}
\mathbf{r}_i(t+\Delta t) = \mathbf{r}_i(t) + \mathbf{v}_i(t)\Delta t
\label{Eq:str1}
\end{equation}
where $\Delta t$ affects the mean free path and the fluid viscosity.
Before the particles perform the collision step they are sorted into cubic cells of length $a_0$ (Fig.~\ref{Fig:1}(b)),
and in the collision step all particles in a cell exchange momentum while the total momentum in the cell is conserved.
Their velocities are updated to
\begin{equation}
\mathbf{v}_i(t+\Delta t) = \mathbf{u}_\xi(t) + \boldsymbol{\Xi}(\mathbf{r}_j(t),\mathbf{v}_j(t))
\label{Eq:coll000}
\end{equation}
where  $\mathbf{u}_\xi(t) $ is the mean velocity in cell $\xi$, and $ \boldsymbol{\Xi}(\mathbf{r}_j(t),\mathbf{v}_j(t)) $
is the collision operator which in general depends on all particle positions and velocities in a cell.
It locally conserves momentum, and depending on the simulated ensemble, it either locally conserves energy or uses a thermostat to keep the temperature constant. 
The dynamics of the fluid particle velocities then determines the spatiotemporal evolution of the coarse-grained fluid velocity field which strongly fluctuates due to the intrinsic particle-based description.

\subsection{Coupling of microswimmers to the fluid}
Typically fluid particles and immersed objects such as microswimmers move with different time steps.
While the fluid particles change their positions and velocities in relatively large steps $\Delta t$,
immersed objects update their position -- and conformation for flexible objects -- at smaller time steps $\delta t$ in order to resolve the molecular dynamics (MD).
In these hybrid MPCD-MD simulations in general there exist two ways how to couple immersed objects to the MPCD fluid.

\begin{figure}[bth]
\begin{center}
\includegraphics[width=7.4cm]{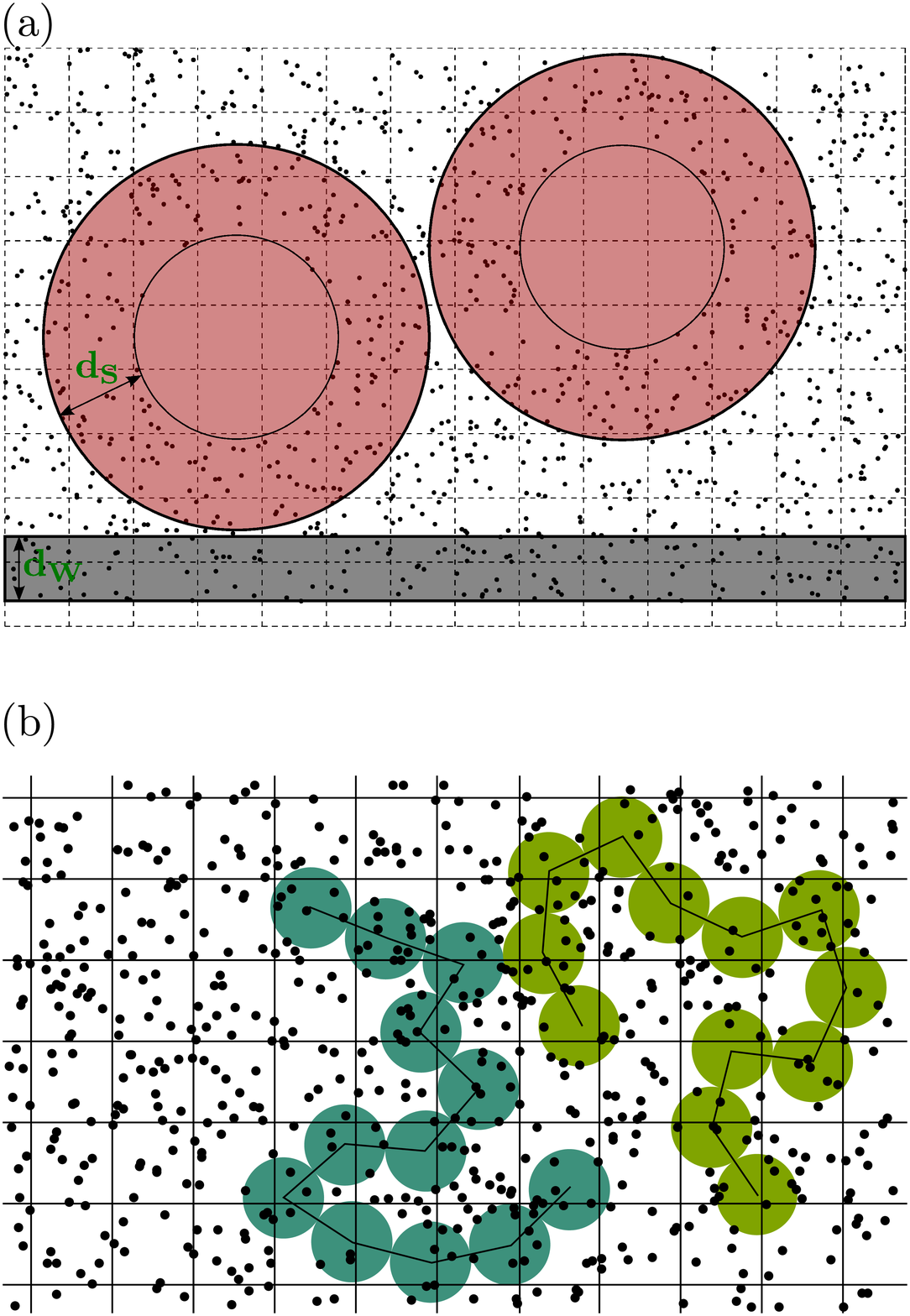}
\end{center}
\caption{Coupling between the MPCD fluid and embedded objects. (a) 2D sketch of  collision step.
\change{Here immersed solid spheres} are shown in red and a wall in grey.
Small dots in the white region indicate the  fluid particles
and dots in the \change{spheres} and in the wall are \textit{virtual} particles,
which are located in a thin layer of thickness $d_W=a_0$ in the wall
and in a thin shell of thickness $d_S = \sqrt{3}a_0$ in the \change{spheres}.
The dashed lines show the collision cell grid (from Ref.~\cite{Zottl2018}).
(b) \change{2D sketch of model polymers in  MPCD fluid:} polymers (light and dark green), MPCD fluid particles (black dots). The rectangular collision cell grid \change{is} shifted randomly in every time step.
 }
\label{Fig:2}
\end{figure}

First, large solid objects are usually not allowed to be penetrated by fluid particles.
Momentum and angular momentum between the objects and the fluid are exchanged in the streaming step. When a fluid particle hits an object it  is bounced back and the local momentum difference is transferred to the object and also modify the angular momentum of the object \cite{Zottl2018}.
This procedure ensures translation and rotation of the objects.
In addition, in order to ensure a no-slip boundary condition momentum and angular momentum can also be exchanged  in the collision step by using so-called \emph{virtual} fluid particles placed in a layer inside the objects (Fig.~\ref{Fig:2}(a)).
These virtual particles are randomly placed each time step and participate in the collision step. The local momentum transfer to the virtual particles then \change{contributes} to the change of momentum and angular momentum of the object.

Second, slender and flexible objects such as flagella, cilia and polymers can be coupled to the fluid solely in the collision step.
This can be achieved  when a filament -- or a membrane -- is discretized by spheres connected by springs, and may be connected by other two- or more body forces such as bending or twist.
Fig.~\ref{Fig:2}(b) shows a 2D sketch of two short filaments suspended in an MPCD fluid.
While the monomers have a certain size $\sigma$ in order to prevent overlap of the filaments, fluid particles are allowed to penetrate the monomers.
With this respect the monomers are treated as pointlike particles, which exchange momentum with the fluid particles only in the collision step.
This is achieved by assigning a constant mass $m_p$ to each of the monomers and which is typically equivalent to the mass of fluid which occupies one collision cell \cite{Gompper2009}.
In this sense fluctuations and hydrodynamic interactions between monomers are included efficiently.
While this method was first used to study polymers in solution \cite{Malevanets2000}, it can also be applied to filaments of flexible microswimmers.

\section{Simulation of deformable microswimmers}

\subsection{Simulations of sperm cells and elastic filaments}
\begin{figure}[bth]
\begin{center}
\includegraphics[width=9cm]{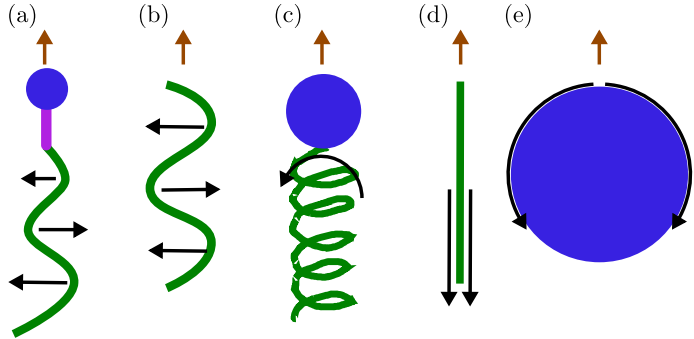}
\end{center}
\caption{Sketch of different microswimmers studied with MPCD. (a) Model sperm cell consisting of a solid head, an attached flexible hook, and a beating elastic tail.
  (b) beating elastic filament. (c) Model bacterium consisting of a solid head with attached helical flagella driven by a motor torque connecting cell body and flagellum. (d) Active rod driven by tangential forces which are counterbalanced by forces on the fluid. (e) Slip-surface driven microswimmers such as squirmers or phoretically driven artificial microswimmers.
 }
\label{Fig:3}
\end{figure}

Using the method of MPCD to model the motion of microswimmers has first been applied to the dynamics of flexible swimmers such as sperm cells
which consist of a spherical head, a hook and a flexible, actively deforming tail (Fig.~\ref{Fig:3}(a)).
Elgeti and coworkers coupled a periodically actuated model sperm to the MPCD fluid in order to study the 3D motion of swimming sperm cells in a background fluid 
\cite{Elgeti2006,Elgeti2008}. \change{An even simpler}  model in 2D has been used to study \change{their} cooperative motion, and showed the importance of hydrodynamic interactions in sperm synchronization and clustering \cite{Yang2008}, and the importance of  hydrodynamic interaction between sperm cells and nearby solid surfaces \cite{Elgeti2010}.
Recent work demonstrated how sperm can be sorted in microfluidic chambers including an array of periodic pillars \cite{Chinnasamy2018}.
Model sperm swimming in modulated channels can be guided by optimizing the channel geometry \cite{Rode2019}.

It is known that flexible waving sheets and filaments can swim force-free without an attached head \cite{Taylor1951,Lauga2009a,Elgeti2015} (Fig.~\ref{Fig:3}(b)).
In 2D the simplest implementation is to apply a bending wave on a one-dimensional filament \cite{Yang2010,Munch2016}.
The large scale collective properties of such waving filaments  show cluster formation and collective swarming \cite{Yang2010}.
When breaking the left-right symmetry of these filaments they are able to swim in circles, similar as observed experimentally for swimming sperm cells near solid surfaces.
The hydrodynamic interactions then determine the formation of vortex arrays in collectively swimming filaments \cite{Yang2014}, in accordance with experimental observations \cite{Riedel2005}.
The dynamics of a single filament in confinement and periodic pillar arrays show a large speed enhancement at certain obstacle configurations, where both steric and hydrodynamic interactions are included \cite{Munch2016}.

Microswimmers such as \change{{\it Paramecia}}  swim by synchronously beating thousands of cilia attached at their cell body.
A large-scale implementation of periodic arrays of cilia shows that hydrodynamic interactions between nearby cilia determine their synchronization  and the formation of metachronal waves \cite{Elgeti2013} which enable ciliates to swim.

Other implementations of flexible microswimmers in MPCD are based on classical simple model swimmers.
First, it is known that a rigid one-hinge swimmer consisting of two rigid arms cannot swim at low Reynolds number because it can only perform time-reversal, reciprocal motion \cite{Purcell1977}.
This time-reversibility can be broken by using finite Reynolds numbers, non-Newtonian fluids \cite{Qiu2014}, or  flexible arms, as recently demonstrated with 2D MPCD simulations \cite{Choudhary2018}.
Second, a very simple model microswimmer which is able to swim is the so-called three-sphere swimmer
which periodically deforms the length of two arms connecting three aligned spheres in a non-reciprocal manner  \cite{Najafi2004}.
This model swimmer has been implemented in the MPCD fluid and extended to simulate the motion of $N$-sphere swimmers, and to the motion of spheres which are not aligned \cite{Earl2007}.

\subsection{Simulation of swimming bacteria}
Many bacteria are able to swim by rotating helical flagella attached to the cell body \cite{Lauga2016} (Fig.~\ref{Fig:3}(c)).
The hydrodynamic interaction between several flagella and their flexibility allows bundle formation and synchronous rotation.
In MPCD the hydrodynamic synchronization of anchored rotating flagella has been studied by discretizing a helix using beads connected by springs and including a bending and torsion potential, in order to include flexibility.
These model flagella are then driven by a torque at the base, similar as for swimming bacteria, and hydrodynamic flagella synchronization can be studied by coupling to the MPCD fluid \cite{Reigh2012,Reigh2013}.
The flow field of an anchored bacterium consisting of a fixed head and an attached rotating flagellum can be determined with MPCD, and it can be shown that the resulting force monopole flow field is mainly determined by the rotating flagellum, and only slightly modified by the presence of the head \cite{Balin2017}.

Hu {\it et al.\ }constructed a more refined mechanical model of bacterial flagella in order to study the bundle formation and swimming of multiflagellated {\it E.~coli} bacteria \cite{Hu2015}.
They were able to  determine  the influence of the specific flagella bundle on the bacterium speed, and to precisely measure the flow field.
It is known that swimming {\it E.~coli} bacteria close to solid surfaces swim in circles, due to the counterrotation of head and flagella tail \cite{DiLuzio2005,Lauga2006}.
Interestingly, using MPCD it can be shown that by gradually increasing the surface slip the shape of the circles is modified and eventually the sense of rotation is inverted \cite{Hu2015a} as known for stress-free surfaces \cite{DiLeonardo2011}.

\begin{figure}[bth]
\begin{center}
\includegraphics[width=11cm]{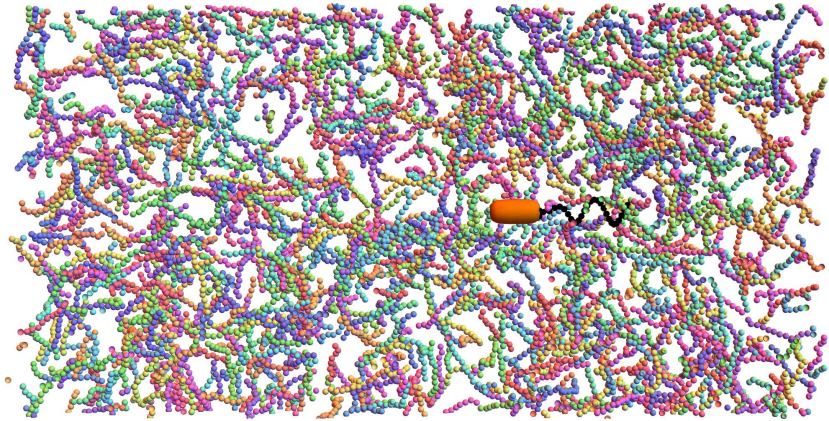}
\end{center}
\caption{Sketch of a model bacterium swimming in a polymer solution consisting of bead-polymers and the background MPCD fluid (not shown).
 }
\label{Fig:4}
\end{figure}

We have recently studied the dynamics of swimming model bacteria in \change{explicitly} modeled polymer solutions, by coupling both the dynamics of the bacterium and the bead-spring polymers to the  background MPCD fluid \cite{Zottl2019} (Fig.~\ref{Fig:4}).
This allowed us to \change{precisely} measure the distribution and conformation of the polymers around the bacterium, and to measure the flow fields for a broad range of polymer densities and polymer types.
The data from the MPCD simulations can then be used to show that a non-uniform distribution of polymers around the bacterium, and in particular around the flagella, leads to local viscosity gradients.
This allowed us to demonstrate that an effective enhanced translation-rotation coupling counterintuitively increase the speed of the bacterium, although the viscosity of the fluids increase with the addition of polymers \cite{Zottl2019}. In contrast, passive driven spheres, rods and ellipsoids are never faster than in water, which we demonstrated recently using MPCD \cite{Zottl2019b}.

\subsection{Swimming by active deformation of a flexible cell body}
The most complex  microswimmers so far implemented with the MPCD method are so-called {\it trypanosomes}
which are highly deformable pathogens which cause the sleeping sickness. \change{They are able to deform} their cell body by the periodic deformation of a firmly attached flagellum which leads to nonreciprocal cell body deformations \change{which allows} them to swim \cite{Babu2012,Babu2012a,Heddergott2012,Alizadehrad2015}.

First the dependence of the swimming speed on the mechanical properties of the organism has been demonstrated  \cite{Babu2012,Babu2012a,Heddergott2012},
and later a refined model has been used to quantitatively match the complex swimming \change{patterns} of experimentally studied swimming {\it African trypanosomes} \cite{Alizadehrad2015}.

\section{Simulation of rigid microswimmers driven by surface flows}

\subsection{Squirmer model}
The squirmer is a simple model microswimmer which originally has been used to model the locomotion and hydrodynamics of swimming ciliated microorganisms such as {\it Volvox} or \change{{\it Paramecia}}.
Cilia at the cell surface of biological microswimmers induce an effective slip velocity close to their cell surface which propels the cell forward \cite{Brennen1977}.
In the simplest squirmer model this slip velocity is  static and axisymmetric at the surface of a spherical particle (Fig.~\ref{Fig:3}(e)), and analytic solutions for the flow fields and the swimming speed exist \cite{Lighthill1952,Blake1971a}.
By tuning a single parameter pushers, pullers, and neutral squirmers can be simulated. Recently  the squirmer became a very \change{successful} model for understanding the hydrodynamic effects in spherical microswimmer suspensions.
The details of the squirmer model and the implementation with MPCD can be found in Ref.~\cite{Zottl2018}.

The squirmer has first been coupled to the MPCD fluid by Downton and Stark \cite{Downton2009a} who were able to approximately recover the analytic bulk flow fields, and to determine the influence of Brownian motion on the dynamics of the microswimmer, which performs a \change{persistent} random walk.
Later the rotational noise has been  captured very accurately by using an angular-momentum conserving version of the MPCD method \cite{Goetze2010} in order to study the stochastic pairwise interaction of two nearby swimming squirmers.

The squirmer can also be used to study the hydrodynamic interactions of  spherical microswimmers with bounding surfaces  \cite{Schaar2015}.
It can be shown that the interplay between near-field hydrodynamic interactions and thermal noise determines the time microswimmers stay at surfaces after collision.
When a squirmer is confined between two parallel plates in a narrow slit, pushers, pullers and neutral swimmers move very differently, again due to hydrodynamic near-field interactions \cite{Zottl2018}, which can be \change{explained} by an analytic model in the noise-free limit (Supplementary Information of Ref.~\cite{Zottl2014}).
While the pairwise swimming of  squirmers in bulk is unstable, bounding surfaces can stabilize the pair-formation of spheroidal squirmers \cite{Theers2016}.

Studying the dynamics of microswimmers such as bacteria in microfluidic devices \change{enables} to study active transport processes under external flow conditions.
Interestingly, when hydrodynamic interactions between microswimmers and bounding walls are neglected, microswimmers do not cluster in a microfluidic channel under flow  but they are \change{homogeneously} distributed.
This changes for pushers and pullers which hydrodynamically interact with the channel walls, as can be demonstrated with MPCD  simulations. Squirmers  either cluster in the center of the channel or near the walls (pullers), or perform stable van-der-Pol-like oscillations in the channel  \cite{Zottl2012a}.

Understanding the collective motion of microswimmers has become an important topic in the field of active matter physics \cite{Zoettl2016,Bechinger2016}.
Typical model systems which neglect hydrodynamic interactions are self-propelled Brownian spheres or rods.
Hydrodynamically interacting squirmers
is a simple model system to study the importance of hydrodynamic interactions in
collectively moving microswimmers.
We studied the dynamics of multiple squirmers confined in a quasi-2D environment using MPCD \cite{Zottl2014} and found that the type of the created flow field plays an important role in the formation of clusters and in motility-induced phase separation.
Later we performed  large-scale simulations in order to identify the role of the hydrodynamic pressure in the phase separation process \cite{Blaschke2016}.
The influence  of fluid compressibility and squirmer elongation on the collective motion in a monolayer has been demonstrated in Ref.~\cite{Theers2018}.

Microswimmers which are not neutrally buoyant but, for example, heavier than water sink due to gravity, which induces external forces and Stokeslet flow fields.
The competition between the sinking of heavy squirmers and the proximity of a nearby wall have been studied recently using MPCD which uncovered complex hoovering motion near walls \cite{Ruhle2018}.
The collective sedimentation of squirmers under gravity shows strong convective currents and non-trivial sedimentation profiles \cite{Kuhr2017}.
In the case of sufficiently low density and strong sedimentation, a monolayer of sedimented squirmers can be formed at a surface, which shows a variety of dynamic states, including a hydrodynamic Wigner fluid \cite{Kuhr2019}.
When the squirmers are not only heavy, but in addition bottom heavy due to an asymmetric mass distribution, the emergent cooperative behavior shows the formation of plums, convection roles and inverted sedimentation \cite{Ruhle2020}.

A recent study showed that the presence of polymers can enhance the rotational diffusion of squirmers \cite{Qi2020}, similar as observed experimentally for a spherical artificial microswimmer moving in viscoelastic fluids \cite{Gomez-Solano2016}.

\subsection{Self-propelled rods and arrays of spheres}
A simple model of a self-propelled rod which generates  hydrodynamic flows around it has been implemented in MPCD in order to study its dynamics close to surfaces \cite{Elgeti2009}.
These rods are driven by a tangential propulsion force (Fig.~\ref{Fig:3}(d)) which is compensated by a force on the fluid, to ensure a force-free system.
Recently the dynamics and distribution of active rods in wall-bounded simple shear flow has been analyzed with a similar model \cite{Anand2019}.

Schwarzendahl and Mazza implemented asymmetric microswimmers consisting of a rigidly connected large and small sphere, which are able to move by imposing force dipoles on the surrounding fluid \cite{Schwarzendahl2018,Schwarzendahl2019,Schwarzendahl2019a}.
They are able to simulate both pusher and puller microswimmers and the implications on their collective motion.

Active dumbbells and the resulting force dipole fields have also been included in order to study the enhanced diffusion of a tracer particle in this active bath \cite{Dennison2017}.

\subsection{Phoretic active particles}
Artificial rigid microswimmers are typically driven by self-phoretic mechanisms such as self-diffusiophoresis, self-thermophoresis or self-electrophoresis \cite{Zoettl2016}.
These mechanisms lead to the creation of local field gradients and fluid flows tangential to the surface, which drive the particle forward.

Reactive multiparticle collision dynamics (R-MPCD) is a \change{simulation} method which is able to simulate the backcoupling between chemical concentration gradients and fluid flow using an explicit particle-based description \cite{Tucci2004}.
It combines the usual MPCD simulation technique with the ability to perform chemical reactions.
Typical simulated microswimmer geometries are either Janus spheres \cite{DeBuyl2013,Yang2014,Huang2016,Huang2017,Huang2018a,Huang2019} or Janus sphere dimers \cite{Rueckner2007,Tao2008a,Tao2009a,Tao2010,Valadares2010,Thakur2011,Thakur2012,Yang2011a,Yang2014}.
The leading order flow field of an active sphere-dimer is a force dipole, as shown in simulations \cite{Yang2014d,Reigh2015} and by an analytic calculation \cite{Reigh2015}.
This can be extended to the case where the catalytically active and inert sphere are not connected but free to move together \cite{Reigh2018} which can lead to the active self-assembly of a microswimmer consisting of two touching spheres \cite{Yu2018}.
Even very small \change{{\AA}ngstrom}-size motors are able to self-propel, despite of large Brownian noise \cite{Colberg2015,Colberg2015a}.
The formation of transient clusters appear in collectively moving chemically active Janus spheres  \cite{Huang2016}.
Cluster formation can be tuned and enhanced by modifying the particle density, and the size of the catalytic cap \cite{Huang2017}.
More elongated collectively moving active sphere dimers in a monolayer show the formation of persistent clusters \cite{Colberg2017}.
When active sphere dimers are pinned on a 2D layer but free to rotate, strong orientational correlations exist, mainly induced by chemical interactions  \cite{Robertson2018}.
If the background medium is chemically inhomogeneous, chemically active microswimmers are able to react to local gradients, or even follow chemical patterns \cite{Chen2018}.
Time-periodic oscillations of the chemical background medium leads to a periodic dispersion-aggregation transition of active nano-dimers \cite{Chen2019}.
The combined influence of chemical patterns and external flow can lead to a chemically induced drift in the channel \cite{Deprez2017}.
Interestingly, when spherical symmetry is broken for active Janus particles, active rotational motion can be induced \cite{deBuyl2019,Reigh2020}.
When several inert and chemically active spheres are connected by flexible bonds, the explicit simulation of these active filaments shows self-propulsion, spontaneous bending and complex shape deformations \cite{Sarkar2016,Sarkar2017}.
By using a simple viscoelastic version of the MPCD fluid it could be shown that self-propelling dimers can enhance their speed substantially \cite{Sahoo2019}.

Self-thermophoretic microswimmers are able to self-generate local temperature gradients around their body which drive them forward \cite{Yang2011a,Yang2014d,Wagner2017,Wagner2019}, or react in thermal gradients and show phototaxis \cite{Yu2019}.
The collective motion of chemically and hydrodynamically interacting thermophoretic sphere dimers shows collective swarming behavior \cite{Wagner2017}.

\section{Conclusion and Perspective}
Here we have reviewed how different classes of microswimmers can be simulated using MPCD, which \change{enables} to identify the influence of hydrodynamic flows on the motion of single or collectively moving microswimmers.
Since the fluid can be modeled \change{explicitly}, complex and time-varying swimming shapes are possible to simulate, as long as the Reynolds number is kept sufficiently small.
Furthermore, by including chemical reactions or temperature gradients, several self-phoretic microswimmers can be simulated.

While a lot of progress has been made in the locomotion of microswimmers in simple fluids, the swimming patterns and strategies of microswimmers in complex fluids and non-trivial environments is a very timely field of research. We expect that the use of MPCD will be used in the near future to study such situations, due to the possibility of solving multi-component and multi-scale systems, such as the combined dynamics of colloids and microswimmers in the presence of \change{explicitly} modeled polymers and background fluid, which just started \cite{Zottl2019,Zottl2019b,Qi2020}.

\change{Furthermore, we expect that the accurate simulation of large-scale hydrodynamic flows created by cooperatively moving microswimmers will become more and more important in the future, in order to understand the importance of the specific microswimmer properties, such as shape and near-field interactions, on the collective dynamics. In MPCD the system size is typically the limiting factor when simulating large-scale dynamics. Hence we expect that massively parallelized systems using modern supercomputer systems will become increasingly important. First implementations so far used multi-CPU parallel computing based on Message Passage Interface  (MPI) \cite{Blaschke2016} or Graphics Processing Units (GPU)  \cite{Theers2016}. In addition we expect hybrid and advanced multiscale methods to be used to bridge different length scales in the system.
Examples are hybrid MD-MPCD fluids separated by a buffer zone \cite{Alekseeva2016}, or hybrid methods combining Lattice Boltzmann (LB) and MPCD fluid domains, again coupled by a buffer zone \cite{Montessori2020}. These methods could become a powerful alternative to previously developed methods which couple large-scale  particle dynamics to continuum hydrodynamic fields (see e.g.\ Refs.~\cite{Saintillan2007,Thampi2016,Li2019a}). }


\addcontentsline{toc}{chapter}{Acknowledgment}
\section*{Acknowledgment}
The author acknowledges funding from the Austrian Science Fund (FWF) through a Lise-Meitner Fellowship (Grant No.~M 2458-N36).

\addcontentsline{toc}{chapter}{References}


\newpage


\begin{thebibliography}{100}

\bibitem{Marchetti2013}
Marchetti M~C, Joanny J~F, Ramaswamy S, Liverpool T~B, Prost J, Rao M and Simha
  R~A 2013 \emph{Rev. Mod. Phys.} \textbf{85} 1143

\bibitem{Vicsek2012}
Vicsek T and Zafeiris A 2012 \emph{Phys. Rep.} \textbf{517} 71

\bibitem{Purcell1977}
Purcell E~M 1977 \emph{Am. J. Phys.} \textbf{45} 3

\bibitem{Lauga2009a}
Lauga E and Powers T~R 2009 \emph{Rep. Prog. Phys.} \textbf{72} 096601

\bibitem{Elgeti2015}
Elgeti J, Winkler R~G and Gompper G 2015 \emph{Rep. Prog. Phys.} \textbf{78}
  056601

\bibitem{Zoettl2016}
Z{\"{o}}ttl A and Stark H 2016 \emph{J. Phys.: Condens. Matter} \textbf{28}
  253001

\bibitem{Prost2015}
Prost J, J{\"{u}}licher F and Joanny J~F 2015 \emph{Nat. Phys.} \textbf{11} 111

\bibitem{Dunweg2009}
D{\"{u}}nweg B and Ladd A~J~C 2009 \emph{Advanced Computer Simulation
  Approaches for Soft Matter Sciences III} 89--166

\bibitem{Espanol2017}
Espa{\~{n}}ol P and Warren P~B 2017 \emph{J. Chem. Phys.} \textbf{146} 150901

\bibitem{Gompper2009}
Gompper G, Ihle T, Kroll D~M and Winkler R~G 2009 \emph{Adv. Poly. Sci.}
  \textbf{221} 1

\bibitem{Kapral2008}
Kapral R 2008 \emph{Adv. Chem. Phys.} \textbf{140} 89

\bibitem{Malevanets1999}
Malevanets A and Kapral R 1999 \emph{J. Chem. Phys.} \textbf{110} 8605

\bibitem{Malevanets2000b}
Malevanets A and Kapral R 2000 \emph{J. Chem. Phys.} \textbf{112} 7260

\bibitem{Zottl2018}
Z{\"{o}}ttl A and Stark H 2018 \emph{Eur. Phys. J. E} \textbf{41} 61

\bibitem{Malevanets2000}
Malevanets a and Yeomans J~M 2000 \emph{Europhys. Lett.} \textbf{52} 231

\bibitem{Elgeti2006}
Elgeti J 2006 \emph{PHD Thesis (Universit{\"{a}}t K{\"{o}}ln)}

\bibitem{Elgeti2008}
Elgeti J and Gompper G 2008 \emph{NIC Symposium 2008} \textbf{39} 53

\bibitem{Yang2008}
Yang Y, Elgeti J and Gompper G 2008 \emph{Phys. Rev. E} \textbf{78} 061903

\bibitem{Elgeti2010}
Elgeti J, Kaupp U~B and Gompper G 2010 \emph{Biophys. J.} \textbf{99} 1018

\bibitem{Chinnasamy2018}
Chinnasamy T, Kingsley J~L, Inci F, Turek P~J, Rosen M~P, Behr B, T{\"{u}}zel E
  and Demirci U 2018 \emph{Adv. Sci.} \textbf{5} 1700531

\bibitem{Rode2019}
Rode S, Elgeti J and Gompper G 2019 \emph{New J. Phys.} \textbf{21} 013016

\bibitem{Taylor1951}
Taylor G 1951 \emph{Proc. Roy. Soc. A} \textbf{209} 447

\bibitem{Yang2010}
Yang Y, Marceau V and Gompper G 2010 \emph{Phys. Rev. E} \textbf{82} 031904

\bibitem{Munch2016}
M{\"{u}}nch J~L, Alizadehrad D, Babu S~B and Stark H 2016 \emph{Soft Matter}
  \textbf{12} 7350

\bibitem{Yang2014}
Yang Y, Qiu F and Gompper G 2014 \emph{Phys. Rev. E} \textbf{89} 012720

\bibitem{Riedel2005}
Riedel I~H, Kruse K and Howard J 2005 \emph{Science} \textbf{309} 300

\bibitem{Elgeti2013}
Elgeti J and Gompper G 2013 \emph{Europhys. Lett.} \textbf{101} 48003

\bibitem{Qiu2014}
Qiu T, Lee T~C, Mark A~G, Morozov K~I, M{\"{u}}nster R, Mierka O, Turek S,
  Leshansky A~M and Fischer P 2014 \emph{Nat. Commun.} \textbf{5} 5119

\bibitem{Choudhary2018}
Choudhary P, Mandal S and Babu S~B 2018 \emph{J. Phys. Commun.} \textbf{2}
  025009

\bibitem{Najafi2004}
Najafi A and Golestanian R 2004 \emph{Phys. Rev. E} \textbf{69} 062901

\bibitem{Earl2007}
Earl D~J, Pooley C~M, Ryder J~F, Bredberg I and Yeomans J~M 2007 \emph{J. Chem.
  Phys.} \textbf{126} 064703

\bibitem{Lauga2016}
Lauga E 2016 \emph{Annu. Rev. Fluid Mech.} \textbf{48} 105

\bibitem{Reigh2012}
Reigh S~Y, Winkler R~G and Gompper G 2012 \emph{Soft Matter} \textbf{8} 4363

\bibitem{Reigh2013}
Reigh S~Y, Winkler R~G and Gompper G 2013 \emph{PloS one} \textbf{8} e70868

\bibitem{Balin2017}
Balin A~K, Z{\"{o}}ttl A, Yeomans J~M and Shendruk T~N 2017 \emph{Phys. Rev.
  Fluids} \textbf{2} 113102

\bibitem{Hu2015}
Hu J, Yang M, Gompper G and Winkler R~G 2015 \emph{Soft Matter} \textbf{11}
  7867

\bibitem{DiLuzio2005}
DiLuzio W~R, Turner L, Mayer M, Garstecki P, Weibel D~B, Berg H~C and
  Whitesides G~M 2005 \emph{Nature} \textbf{435} 1271

\bibitem{Lauga2006}
Lauga E, DiLuzio W~R, Whitesides G~M and Stone H~A 2006 \emph{Biophys. J.}
  \textbf{90} 400

\bibitem{Hu2015a}
Hu J, Wysocki A, Winkler R~G and Gompper G 2015 \emph{Sci. Rep.} \textbf{5}
  9586

\bibitem{DiLeonardo2011}
{Di Leonardo} R, Dell'Arciprete D, Angelani L and Iebba V 2011 \emph{Phys. Rev.
  Lett.} \textbf{106} 038101

\bibitem{Zottl2019}
Z{\"{o}}ttl A and Yeomans J~M 2019 \emph{Nat. Phys.} \textbf{15} 554

\bibitem{Zottl2019b}
Z{\"{o}}ttl A and Yeomans J~M 2019 \emph{J. Phys.: Condens. Matter} \textbf{31}
  234001

\bibitem{Babu2012}
Babu S~B and Stark H 2012 \emph{New J. Phys.} \textbf{14} 085012

\bibitem{Babu2012a}
Babu S~B, Schmeltzer C and Stark H 2012 \emph{Nature-Inspired Fluid Mechanics}
  \textbf{119} 25

\bibitem{Heddergott2012}
Heddergott N, Kr{\"{u}}ger T, Babu S~B, Wei A, Stellamanns E, Uppaluri S, Pfohl
  T, Stark H and Engstler M 2012 \emph{PLoS Pathog.} \textbf{8} e1003023.

\bibitem{Alizadehrad2015}
Alizadehrad D, Kr{\"{u}}ger T, Engstler M and Stark H 2015 \emph{PLoS Comput.
  Biol.} \textbf{11} e1003967.

\bibitem{Brennen1977}
Brennen C and Winet H 1977 \emph{Ann. Rev. Fluid Mech.} \textbf{9} 339

\bibitem{Lighthill1952}
Lighthill J~M 1952 \emph{Commun. Pure Appl. Math.} \textbf{5} 109

\bibitem{Blake1971a}
Blake J~R 1971 \emph{J. Fluid Mech.} \textbf{46} 199

\bibitem{Downton2009a}
Downton M~T and Stark H 2009 \emph{J. Phys.: Condens. Matter} \textbf{21}
  204101

\bibitem{Goetze2010}
G{\"{o}}tze I~O and Gompper G 2010 \emph{Phys. Rev. E} \textbf{82} 041921

\bibitem{Schaar2015}
Schaar K, Z{\"{o}}ttl A and Stark H 2015 \emph{Phys. Rev. Lett.} \textbf{115}
  038101

\bibitem{Zottl2014}
Z{\"{o}}ttl A and Stark H 2014 \emph{Phys. Rev. Lett.} \textbf{112} 118101

\bibitem{Theers2016}
Theers M, Westphal E, Gompper G and Winkler R~G 2016 \emph{Soft Matter}
  \textbf{12} 7372

\bibitem{Zottl2012a}
Z{\"{o}}ttl A and Stark H 2012 \emph{Phys. Rev. Lett.} \textbf{108} 218104

\bibitem{Bechinger2016}
Bechinger C, {Di Leonardo} R, L{\"{o}}wen H, Reichhardt C, Volpe G and Volpe G
  2016 \emph{Rev. Mod. Phys.} \textbf{88} 045006

\bibitem{Blaschke2016}
Blaschke J, Maurer M, Menon K, Z{\"{o}}ttl A and Stark H 2016 \emph{Soft
  Matter} \textbf{12} 9821

\bibitem{Theers2018}
Theers M, Westphal E, Qi K, Winkler R~G and Gompper G 2018 \emph{Soft Matter}
  \textbf{14} 8590

\bibitem{Ruhle2018}
R{\"{u}}hle F, Blaschke J, Kuhr J~T and Stark H 2018 \emph{New J. Phys.}
  \textbf{20} 025003

\bibitem{Kuhr2017}
Kuhr J~T, Blaschke J, R{\"{u}}hle F and Stark H 2017 \emph{Soft Matter}
  \textbf{13} 7548

\bibitem{Kuhr2019}
Kuhr J~T, R{\"{u}}hle F and Stark H 2019 \emph{Soft Matter} \textbf{15} 5685

\bibitem{Ruhle2020}
R{\"{u}}hle F and Stark H 2020 \emph{Arxiv preprint} arXiv:2020.04323

\bibitem{Qi2020}
Qi K, Westphal E, Gompper G and Winkler R~G 2020 \emph{Phys. Rev. Lett.}
  \textbf{124} 68001

\bibitem{Gomez-Solano2016}
Gomez-Solano J~R, Blokhuis A and Bechinger C 2016 \emph{Phys. Rev. Lett.}
  \textbf{116} 138301

\bibitem{Elgeti2009}
Elgeti J and Gompper G 2009 \emph{Europhys. Lett.} \textbf{85} 38002

\bibitem{Anand2019}
Anand S~K and Singh S~P 2019 \emph{Soft Matter} \textbf{15} 4008

\bibitem{Schwarzendahl2018}
Schwarzendahl F~J and Mazza M~G 2018 \emph{Soft Matter} \textbf{14} 4666

\bibitem{Schwarzendahl2019}
Schwarzendahl F~J and Mazza M~G 2019 \emph{arXiv preprint} arXiv:1908.10631

\bibitem{Schwarzendahl2019a}
Schwarzendahl F~J and Mazza M~G 2019 \emph{J. Chem. Phys.} \textbf{150} 184902

\bibitem{Dennison2017}
Dennison M, Kapral R and Stark H 2017 \emph{Soft Matter} \textbf{13} 3741

\bibitem{Tucci2004}
Tucci K and Kapral R 2004 \emph{J. Chem. Phys.} \textbf{120} 8262

\bibitem{DeBuyl2013}
de~Buyl P and Kapral R 2013 \emph{Nanoscale} \textbf{5} 1337

\bibitem{Huang2016}
Huang M~J, Schofield J and Kapral R 2016 \emph{Soft Matter} \textbf{12} 5581

\bibitem{Huang2017}
Huang M~J, Schofield J and Kapral R 2017 \emph{New J. Phys.} \textbf{19} 125003

\bibitem{Huang2018a}
Huang M~J, Schofield J, Gaspard P and Kapral R 2018 \emph{J. Chem. Phys.}
  \textbf{149} 024904

\bibitem{Huang2019}
Huang M~J, Schofield J, Gaspard P and Kapral R 2019 \emph{J. Chem. Phys.}
  \textbf{150} 124110

\bibitem{Rueckner2007}
R{\"{u}}ckner G and Kapral R 2007 \emph{Phys. Rev. Lett.} \textbf{98} 150603

\bibitem{Tao2008a}
Tao Y~G and Kapral R 2008 \emph{J. Chem. Phys.} \textbf{128} 164518

\bibitem{Tao2009a}
Tao Y~G and Kapral R 2009 \emph{J. Chem. Phys.} \textbf{131} 024113

\bibitem{Tao2010}
Tao Y~G and Kapral R 2010 \emph{Soft Matter} \textbf{6} 756

\bibitem{Valadares2010}
Valadares L~F, Tao Y~G, Zacharia N~S, Kitaev V, Galembeck F, Kapral R and Ozin
  G~A 2010 \emph{Small} \textbf{6} 565

\bibitem{Thakur2011}
Thakur S and Kapral R 2011 \emph{J. Chem. Phys.} \textbf{135} 024509

\bibitem{Thakur2012}
Thakur S and Kapral R 2012 \emph{Phys. Rev. E} \textbf{85} 026121

\bibitem{Yang2011a}
Yang M and Ripoll M 2011 \emph{Phys. Rev. E} \textbf{84} 061401

\bibitem{Yang2014d}
Yang M, Wysocki A and Ripoll M 2014 \emph{Soft Matter} \textbf{10} 6208

\bibitem{Reigh2015}
Reigh S~Y and Kapral R 2015 \emph{Soft Matter} \textbf{11} 3149

\bibitem{Reigh2018}
Reigh S~Y, Chuphal P, Thakur S and Kapral R 2018 \emph{Soft Matter} \textbf{14}
  6043

\bibitem{Yu2018}
Yu T, Chuphal P, Thakur S, Reigh S~Y, Singh D~P and Fischer P 2018 \emph{Chem.
  Commun.} \textbf{54} 11933

\bibitem{Colberg2015}
Colberg P~H and Kapral R 2015 \emph{J. Chem. Phys.} \textbf{143} 184906

\bibitem{Colberg2015a}
Colberg P~H and Kapral R 2014 \emph{Europhys. Lett.} \textbf{106} 30004

\bibitem{Colberg2017}
Colberg P~H and Kapral R 2017 \emph{J. Chem. Phys.} \textbf{147} 064910

\bibitem{Robertson2018}
Robertson B, Stark H and Kapral R 2018 \emph{Chaos} \textbf{28} 045109

\bibitem{Chen2018}
Chen J~X, Chen Y~G and Kapral R 2018 \emph{Adv. Sci.} \textbf{5} 1800028

\bibitem{Chen2019}
Chen J~X, Zhan S, Qiao L~Y, Ding H~L and Ma Y~Q 2019 \emph{Europhys. Lett.}
  \textbf{125} 26002

\bibitem{Deprez2017}
Deprez L and de~Buyl P 2017 \emph{Soft Matter} \textbf{13} 3532

\bibitem{deBuyl2019}
de~Buyl P 2019 \emph{Phys. Rev. E} \textbf{100} 022603

\bibitem{Reigh2020}
Reigh S~Y, Huang M~J, L{\"{o}}wen H, Lauga E and Kapral R 2020 \emph{Soft
  Matter} \textbf{16} 1236

\bibitem{Sarkar2016}
Sarkar D and Thakur S 2016 \emph{Phys. Rev. E} \textbf{93} 032508

\bibitem{Sarkar2017}
Sarkar D and Thakur S 2017 \emph{J. Chem. Phys.} \textbf{146} 154901

\bibitem{Sahoo2019}
Sahoo S, Singh S~P and Thakur S 2019 \emph{Soft Matter} \textbf{15} 2170

\bibitem{Wagner2017}
Wagner M and Ripoll M 2017 \emph{Europhys. Lett.} \textbf{119} 66007

\bibitem{Wagner2019}
Wagner M and Ripoll M 2019 \emph{Int. J. Mod. Phys. C} \textbf{30} 1941008

\bibitem{Yu2019}
Yu N, Lou X, Chen K and Yang M 2019 \emph{Soft Matter} \textbf{15} 408

\bibitem{Alekseeva2016}
Alekseeva U, Winkler R~G and Sutmann G 2016 \emph{J. Comput. Phys.}
  \textbf{314} 14

\bibitem{Montessori2020}
Montessori A, Tiribocchi A, Lauricella M and Succi S 2020 \emph{Arxiv preprint}
  arXiv:2004.00304

\bibitem{Saintillan2007}
Saintillan D and Shelley M~J 2007 \emph{Phys. Rev. Lett.} \textbf{99} 058102

\bibitem{Thampi2016}
Thampi S~P, Doostmohammadi A, Shendruk T~N, Golestanian R and Yeomans J~M 2016
  \emph{Sci. Adv.} \textbf{2} e1501854

\bibitem{Li2019a}
Li H, Shi X, Huang M, Chen X, Xiao M, Liu C, Chat{\'{e}} H and Zhang H~P 2019
  \emph{Proc. Natl. Acad. Sci. (U.S.A.)} \textbf{116} 777

\end{thebibliography}

\end{document}